\documentclass[final,5p,times,twocolumn]{elsarticle} 

\usepackage{lineno,hyperref,graphicx,caption,subcaption,amsmath}
\modulolinenumbers[200]
\journal{Nuclear Instruments and Methods in Physics Research A}

\begin{document}

\begin{frontmatter}

\title{Preliminary results of the Single Event Effect testing for the ULTRASAT sensors}

%% Group authors per affiliation:
%\author{Vlad D. Berlea\fnref{myfootnote}}
%\fntext[myfootnote]{Since 1880.}

%% or include affiliations in footnotes:
\author[mymainaddress,otheraddress]{Vlad D. Berlea}
\author[mymainaddress]{Arooj Asif}
\author[mymainaddress]{Merlin F. Barschke}
\author[mymainaddress,otheraddress]{David Berge}
\author[mymainaddress]{Juan Maria Haces Crespo}
\author[mymainaddress]{Gianluca Giavitto}

\author[mymainaddress]{Shashank Kumar}
\author[mymainaddress]{Andrea Porelli}
\author[mymainaddress]{Nicola de Simone}

\author[mymainaddress]{Jason Watson}
\author[mymainaddress,otheraddress]{Steven Worm}
\author[mymainaddress]{Francesco Zappon}

\author[mysecondaryaddress]{Adi Birman}
\author[mysecondaryaddress]{Shay Alfassi}
\author[mysecondaryaddress]{Amos Fenigstein}

\author[mythirdaddress]{Eli Waxman}
\author[mythirdaddress]{Udi Netzer}
\author[mythirdaddress]{Tuvia Liran}
\author[mythirdaddress]{Ofer Lapid}
\author[myfourthaddress]{Viktor M. Algranatti}
\author[mythirdaddress]{Yossi Shvartzvald}

\address[mymainaddress]{Deutsches Elektronen-Synchrotron DESY, Platanenallee 6, 15738 Zeuthen, Germany}
\address[otheraddress] {Institut f{\"u}r Physik, Humboldt-Universit{\"a}t zu Berlin, Newtonstrasse 15, 12489 Berlin, Germany}
\address[mysecondaryaddress]{Tower Semiconductor, 20 Shaul Amor Avenue, Migdal Haemek, 2310502, Israel}
\address[mythirdaddress]{Department of Particle Physics and Astrophysics, Weizmann Institute of Science, Herzl St 234, Rehovot, Israel}
\address[myfourthaddress]{Israel Space Agency, Tel Aviv, Israel}

\begin{abstract}
ULTRASAT (ULtra-violet TRansient Astronomy SATellite) is a wide-angle space telescope that will perform a deep time-resolved all-sky survey in the near-ultraviolet (NUV) spectrum. The science objectives are the detection of counterparts to short-lived transient astronomical events such as gravitational wave sources and supernovae. The mission is led by the Weizmann Institute of Science and is planned for launch in 2026 in collaboration with the Israeli Space Agency and NASA. DESY will provide the UV camera, composed by the detector assembly located in the telescope focal plane and the remote electronics unit. The camera is composed out of four back-metallized CMOS Image Sensors (CIS) manufactured in the 4T, dual gain Tower process. As part of the radiation qualification of the camera, Single Event Effect (SEE) testing has been performed by irradiating the sensor with heavy ions at the RADEF, Jyvaskyla facility. Preliminary results of both Single Event Upset (SEU) and Single Event Latch-up (SEL) occurrence rate in the sensor are presented. Additionally, an in-orbit SEE rate simulation has been performed in order to gain preliminary knowledge about the expected effect of SEE on the mission.
\end{abstract}

\begin{keyword}
\texttt 4T CMOS Imaging Sensor (CIS), Tower $180$~$nm$, Backside Illumination (BSI), Single Event Effects
%\MSC[2010] 00-01\sep  99-00
\end{keyword}

\end{frontmatter}

\linenumbers

\section{Introduction}

ULTRASAT is a wide-angle space telescope that will perform deep time-resolved all-sky scans in the near-ultraviolet (NUV) band. The science objectives are the detection of counterparts to gravitational wave sources and supernovae \citep{0}.

The UV camera is comprised of four independent CMOS imaging sensors (CIS) with a total active area of $81$~cm$^2$ and $90$~Megapixel \citep{5}. Each individual sensor is built in the $180$ nm Tower feature size process. Each relatively large sensor, with an active area of $4.5$ x $4.5$~cm$^2$ is achieved, through the stitching of several blocks. A backside illumination process is employed to increase the sensor's quantum efficiency.

During the 3-6 years of operation in the geostationary orbit, the sensors will see an elevated proton and heavy ion flux from several sources. This particle flux is expected to lead to both cumulative \citep{4} radiation damage (most importantly in terms of total ionizing damage) and Single Event Effects (SEE). This paper shows first results of the SEE impact on the ULTRASAT sensors.
\if false
\begin{figure}[h]
	\includegraphics[width=.55\columnwidth]
	{ULTRASAT_cam.png}\hfill
	\includegraphics[width=.45\columnwidth]
	{ULTRASAT_sat.png}
	
	\caption{Computer rendering of the ULTRASAT camera design (left) and the ULTRASAT satellite (right). }
	\label{fig:renders}
\end{figure}
\fi
\section{Single Event Effect testing}

The main SEE mechanisms expected for the ULTRASAT sensor are the Single Event Upsets (SEUs) in the on-chip memory circuits and Single Event latch-ups (SELs) in the digital circuitry of the Micro Controller Unit (MCU). The SEU cross-section was estimated by continuously reading and writing the entire memory array. The SEL cross-section was measured with the help of a software switch with a configurable current threshold ($I_{thr}=100$~mA above baseline) and a time over threshold counter. Once ten consecutive samples above the threshold are registered ($5$~Hz), the digital and analog power to the sensor is cycled and an event is counted.

A SEE testing campaign has been performed at room temperature and in air the Jyvaskyla RADEF facility \citep{1}. This test campaign took advantage of the large linear energy transfer (LET) parameter range of the facility's $16.3$~MeV/n cocktail: $13-67$~MeV cm$^2/$mg. The analog and digital power domains were biased at their maximum values, $3.6$ V, while all other lines were grounded.

\section{Experimental results}

\begin{figure}[h]
	\includegraphics[width=1\columnwidth]
	{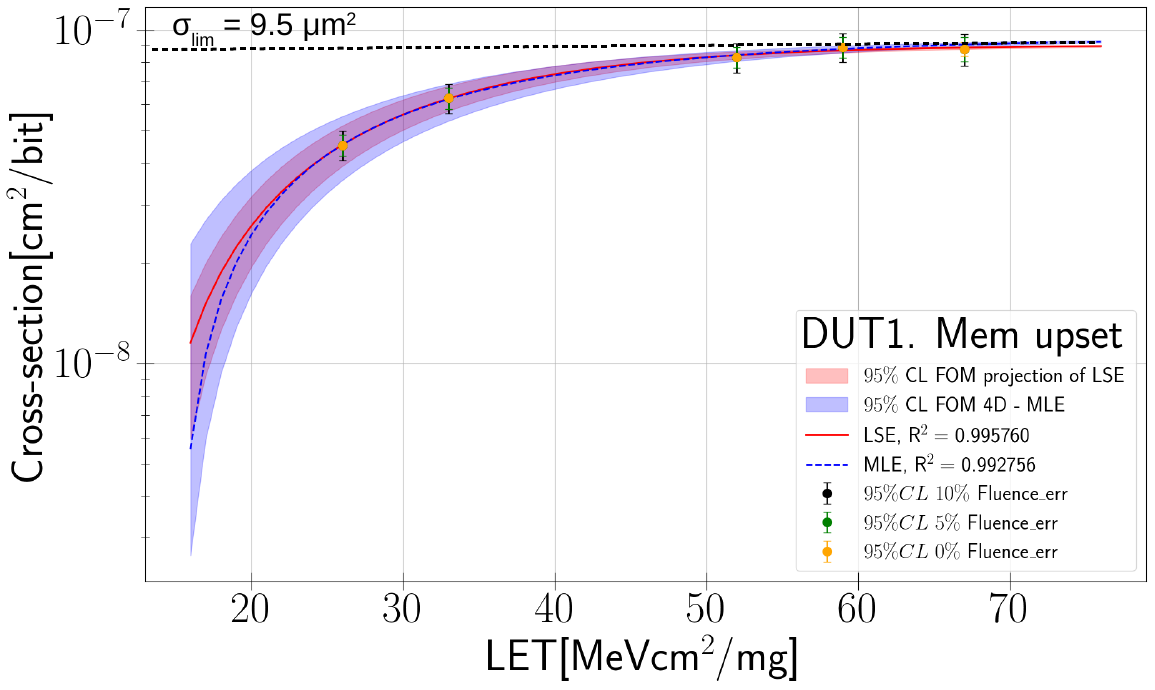}\hfill
	
	\caption{Single event upset cross-section of the ULTRASAT on-chip memory circuit. The data is fitted with the help of two techniques: the Least Squares Estimation (with red) and the Maximum Likelihood Estimation (with blue) showing compatible results.}
	\label{fig:SEU}
\end{figure}

The test campaign showed that both SEUs and SELs are detected in the ULTRASAT sensor for particles with LETs over $13$~MeV cm$^2/$mg. As expected, an increase in the cross-sections of both effects was observed with the increase in particle's LET. The SEU cross-section for several measured LET's can be seen in Figure \ref{fig:SEU} and is fitted by the 4-parameter Weibull function, featuring a good convergence. 

\if false
\begin{align}
\label{eq:Weibull}
    \text{f}(\text{x}) = \sigma_{\mathrm{lim}}(1-\text{e}^{(-(\text{x}-\text{LET}_0)/\text{l})^\text{k}}) 
\end{align}
\fi

\begin{figure}[h]
	\includegraphics[width=1\columnwidth]
	{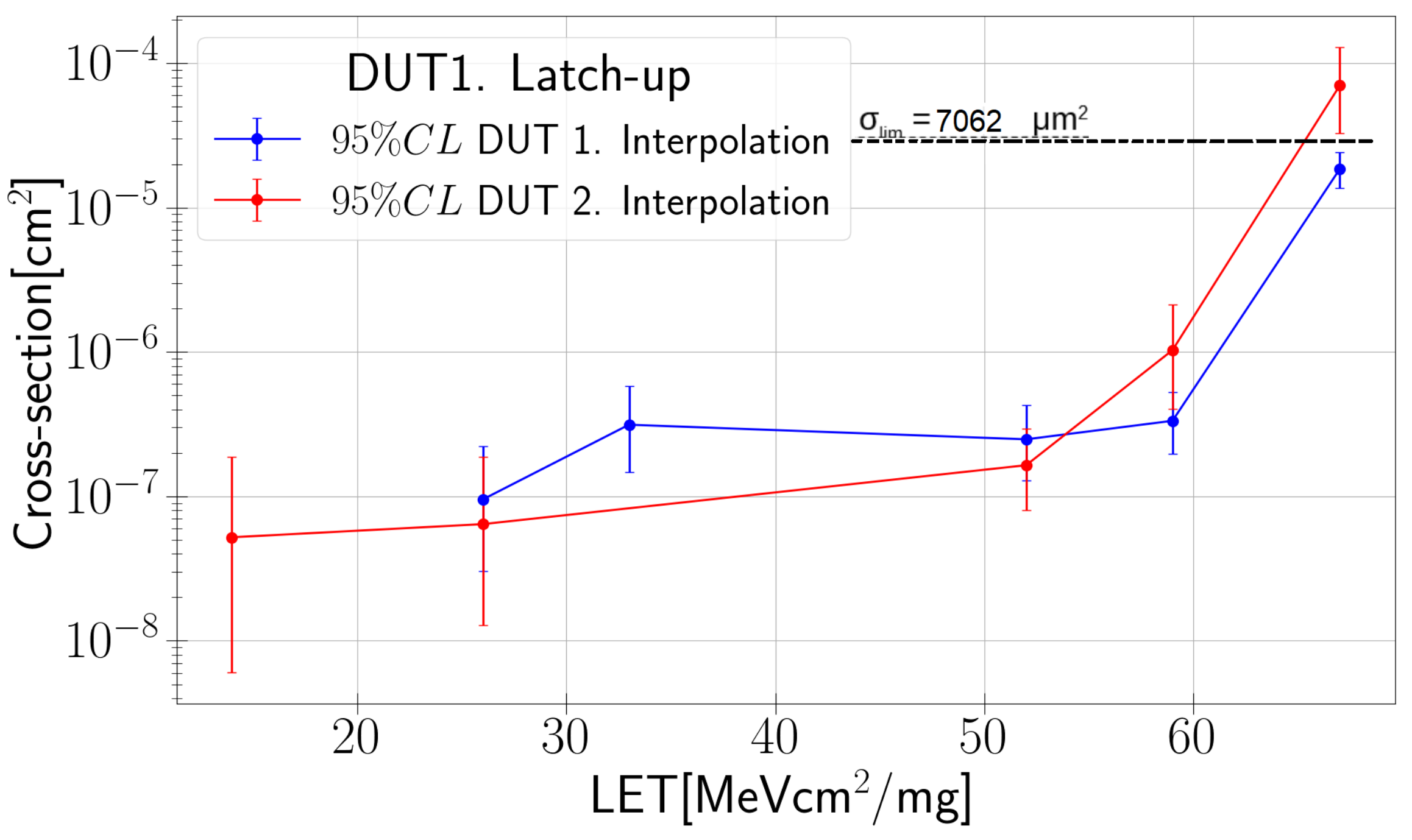}\hfill
	
	\caption{Single event latch-up cross-section of the ULTRASAT sensor for two measured devices under test (DUTs).}
	\label{fig:SEL}
\end{figure}

The experimental curves for the measured SEL for two DUT's are presented in Figure \ref{fig:SEL}. The two curves show comparable results within $2\sigma$, indicating a small inter-sample result variation. In contrast to the showcased SEU data, the SEL results significantly deviate from the Weibull ``turn-on" behaviour. Due to the relatively limited LET range of the experiment, a conclusion on the observed data can not yet be made. However, by investigating the over-currents that were measured in the sensor after a latch-up event a trend in over-current values was observed. Relatively small over-current events ($<250$~mA) were detected only for LETs above $58$~MeV cm$^2/$mg. This fact hints towards the possible existence of two latch-up mechanisms inside the sensor: a ``conventional" large over-current event associated to a global latch of a parasitic CMOS thyristor structure and a small over-current micro-latch event \citep{2} within a smaller division of the circuit.

\section{Flux simulation and in-orbit rate estimation}

\begin{figure}[h]
	\includegraphics[width=1\columnwidth]
	{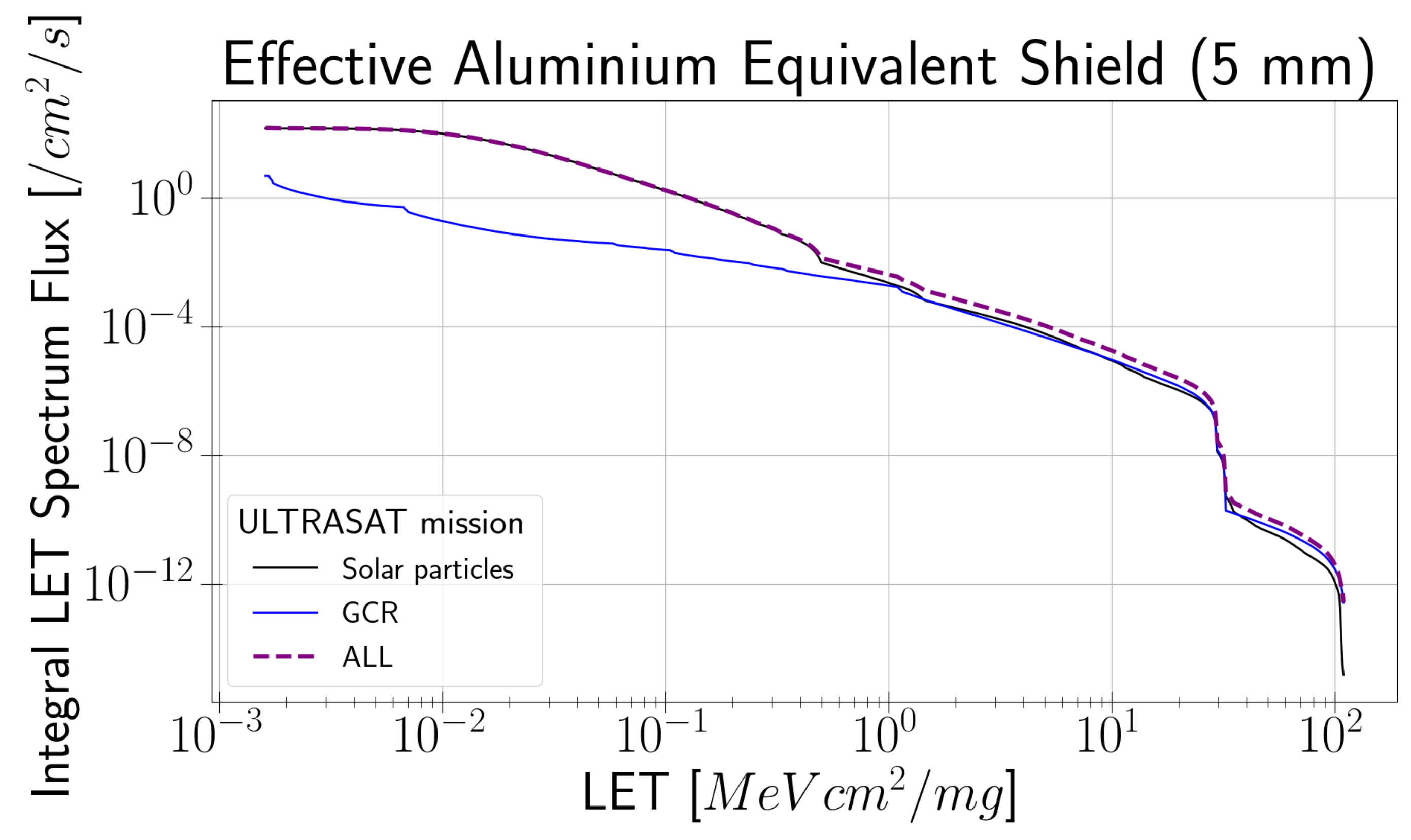}\hfill
	
	\caption{Integral flux in GEO orbit versus LET for the ULTRASAT effective Aluminium equivalent shielding ($5.05$ mm).}
	\label{fig:flux}
\end{figure}

To contextualize the above measurements, a simulation of the expected particle flux for the ULTRASAT mission was performed with the help of the SPENVIS tool \citep{3}. The simulation considers an accurate estimation of the GEO orbit, the launch date of the mission (late 2027), the effective radiation shielding ($\sim 5$~mm) and the various particle species shown in Figure \ref{fig:flux}: trapped proton flux (AP-8 MIN model); trapped electron flux (AE-8 MIN model); solar particle flux (ESP-PSYCHIC model); galactic cosmic ray flux (CREME-96 Solar Min. model). 

Lastly, the expected rate in orbit was derived by integrating the measured cross-section and integral flux rate for each particle track within the sensor (assuming an isotropic particle flux). Preliminary results show an expected SEU rate of $2$ events and an SEL rate of $\sim5\cdot10$$^{-3}$ events across a mission lifetime of $3$ years.

\section{Conclusions and outlook}

The SEE testing of the ULTRASAT sensor has shown that both upsets and latch-ups are detected under the exposure of heavy ionizing particles with LETs in the range of $13-67$~MeV~cm$^2/$~mg. While the upset estimation follows the expected Weibull curve ``turn-on" behavior, unexpected results were obtained for the latch-ups. A possible explanation for these results was postulated to be the existence of two latch-up mechanisms which can be differentiated by their over-current value. While the latch-up occurrence rate is predicted to be relatively small (additionally considering the beneficial effect of the $200$~K camera operation in space), a hardware latch-up detection and power cycling solution is being designed for the on-board spacecraft electronics.

In order to gain a better understanding of the underlying physics effects and achieve a more accurate bounding of the estimated rate, another testing campaign has been scheduled in June 2024 at the SIRAD facility in Padova. This will be achieved by testing the low and high LET ranges of the SEL and SEU cross-sections.

%\section*{References}
\bibliographystyle{elsarticle-num} 
\bibliography{main.bib}

\end{document}